\documentclass{PoS}

\usepackage{verbatim}
\usepackage{float}
\usepackage{amsmath}
\usepackage{amssymb}
\usepackage{graphicx}

\makeatletter

\newcommand {\beq} {\begin{equation}}
\newcommand {\eeq} {\end{equation}}
\newcommand {\beqa}{\begin{eqnarray}}
\newcommand {\eeqa}{\end{eqnarray}}

\title{Spontaneous symmetry breaking induced by complex fermion determinant 
--- yet another success of the complex Langevin method\thanks{KEK-TH/1948}}

\ShortTitle{Spontaneous symmetry breaking induced by complex fermion determinant 
}

\author{\speaker{Yuta Ito}\\
KEK Theory Center, 
High Energy Accelerator Research Organization,\\
1-1 Oho, Tsukuba, Ibaraki 305-0801, Japan\\
        E-mail: \email{yito@post.kek.jp}}

\author{Jun Nishimura\\
KEK Theory Center, 
High Energy Accelerator Research Organization,\\
1-1 Oho, Tsukuba, Ibaraki 305-0801, Japan\\
Graduate University for Advanced Studies (SOKENDAI),\\
1-1 Oho, Tsukuba, Ibaraki 305-0801, Japan\\
        E-mail: \email{jnishi@post.kek.jp}}

\abstract{In many interesting systems, 
the fermion determinant becomes complex 
and its phase plays a crucial role in the determination of the vacuum. 
For instance, in finite density QCD at low temperature and high density, 
exotic fermion condensates are conjectured to form due to such effects. 
When one applies the complex Langevin method to such a complex action system 
naively, one cannot obtain the correct results 
because of the singular-drift problem associated with the appearance 
of small eigenvalues of the Dirac operator. 
Here we propose to avoid this problem
by adding a fermion bilinear term to the action 
and extrapolating 
its coefficient to zero. We test this idea in an 
SO(4)-invariant matrix model with a Gaussian action 
and a complex fermion determinant, 
whose phase is expected to induce the spontaneous breaking 
of the SO(4) symmetry. 
Our results agree well with the previous results 
obtained by the Gaussian expansion method.
}

\FullConference{34th annual International Symposium on Lattice Field Theory\\
		24-30 July 2016\\
		University of Southampton, UK}

\begin{document}
\section{Introduction\label{sec:Introduction}}

Monte Carlo methods are difficult to apply
to a system with a complex action $S$
because of a notorious technical problem known as
the sign problem.
The importance sampling cannot be applied
since the integrand $\exp\left(-S\right)$
of the partition function cannot be regarded as the
probability distribution.
The complex Langevin method (CLM) \cite{Parisi:1984cs,Klauder:1983sp}
is a promising approach to such complex-action systems.
It may be regarded as an extension of the stochastic quantization based on
the Langevin equation, where the dynamical variables of the original system
are complexified and the observables as well as the drift
term are extended holomorphically by analytic continuation.

An important remaining problem of the CLM
is that the method gives wrong results
when the determinant that appears from integrating out fermions takes
values close to zero during the complex Langevin 
simulation \cite{Mollgaard:2013qra}.
A theoretical understanding of this problem
has been given recently \cite{Nishimura:2015pba}, where
it was recognized that the probability distribution of the
complexified variables has to fall off fast enough near the singularities
of the drift term in order to justify the CLM along the line of 
the argument \cite{Aarts:2011ax}. 
See also ref.~\cite{Nagata:2016vkn} for the proposal of
a new criterion for justification based on a refined argument.

In many systems with a complex fermion determinant,
its phase is expected to play a crucial
role in the determination of the vacuum. 
For instance,
in finite density QCD at low temperature and high density, 
the phase is conjectured to induce the formation of
various exotic fermion condensates.
Also 
in the Euclidean version of the type IIB matrix model \cite{Ishibashi:1996xs}
for 10d superstring theory, 
the phase is conjectured to induce 
the SSB of the SO(10) rotational symmetry.

When one applies the CLM to these systems, the singular-drift problem
occurs because of the appearance of eigenvalues of the Dirac operator
close to zero. Here we propose to avoid this problem by deforming
the action with a fermion bilinear term and extrapolating its coefficient
to zero. The fermion bilinear term has to be chosen
carefully in such a way
that the nearly zero eigenvalues of the Dirac operator are avoided
and yet the vacuum of the system is minimally affected.

We test this idea in an SO(4)-symmetric matrix model with a Gaussian
action and a complex fermion determinant, in which the SSB
of SO(4) symmetry is expected to occur due to the phase of the 
determinant \cite{Nishimura:2001sq}. 
When one applies the CLM to this system, the singular-drift problem
is actually severe because the fermionic part of the model is
essentially ``massless''.
Using the idea described above, 
we find that the SO(4) symmetry of the original matrix model is broken
spontaneously down to SO(2).
Moreover, the order parameters thus obtained turn out to be consistent
with the prediction obtained 
by the Gaussian expansion method (GEM) \cite{Nishimura:2004ts}. 
Note that we are able to determine the true vacuum directly without
having to compare the free energy for each vacuum preserving different
amount of rotational symmetry unlike the case with the GEM.
For more details, we refer the readers to our paper \cite{Ito:2016efb}.

The rest of this article is organized as follows. 
In section \ref{sec:The-definition-of},
we define the SO(4)-symmetric matrix model.
In section \ref{sec:Application-of-the}, we explain how we apply
the CLM to the SO(4)-symmetric matrix model. In particular, we deform
the action with a fermion bilinear term, which enables us to investigate
the SSB without suffering from the singular-drift problem. In section
\ref{sec:results}, we present the results of our analysis. In particular,
we extrapolate the deformation parameter to zero, and confirm that
the SSB from SO(4) to SO(2) indeed occurs in this model.
Section \ref{sec:Summary-and-discussion} is devoted to a summary
and discussions.

\section{Brief review of the SO(4)-symmetric matrix model \label{sec:The-definition-of}}

The SO(4)-symmetric matrix model we study is defined
by the partition function \cite{Nishimura:2001sq} 
\begin{equation}
Z=\int dX\,\left(\det D\right)^{N_{\mathrm{f}}}e^{-S_{{\rm b}}} \ ,
\label{part-fn-with-det}
\end{equation}
where the bosonic part of the action is given as 
\begin{eqnarray}
S_{{\rm b}} & = & \frac{1}{2}
N\sum_{\mu=1}^{4}{\rm tr}\,(X_{\mu})^{2}\ .\label{eq:boson_action}
\end{eqnarray}
Here we have introduced $N\times N$ Hermitian matrices $X_{\mu}$
$(\mu=1,\ldots,4)$. 
The Dirac operator $D$ in (\ref{part-fn-with-det}) is represented
by a $2N \times 2N$ matrix
\begin{equation}
D_{i\alpha,j\beta}=\sum_{\mu=1}^{4}(\Gamma_{\mu})_{\alpha\beta}(X_{\mu})_{ij}
\;\;\text{with }\
\Gamma_{\mu}=\begin{cases}
i\,\sigma_{i} & \text{for}\;\mu=i=1,2,3\ ,\\
\mathbf{1}_{2} & \text{for~}\mu=4\  ,
\end{cases}\label{eq:dirac_op}
\end{equation}
where the $2\times2$ matrices $\Gamma_{\mu}$
are the gamma matrices in 4d Euclidean space after Weyl projection,
which are defined by using the Pauli matrices $\sigma_{i}$ ($i=1,2,3$).
The model has an SO(4) symmetry, under which $X_{\mu}$ transforms
as a vector.
The fermion determinant $\det D$ is complex in general.
%

The SO(4) rotational symmetry of the model
is speculated to be
spontaneously broken in the large-$N$ limit with fixed $r=N_{{\rm f}}/N>0$
due to the effect of the phase of the determinant \cite{Nishimura:2001sq}.
In the phase-quenched model, which is defined by omitting the phase
of the fermion determinant, the SSB was shown not to occur by Monte
Carlo simulation \cite{Anagnostopoulos:2011cn}. 
We may therefore say that the SSB should be
induced by the phase of the fermion determinant. 
Below we restrict ourselves to the $r=1$ case, 
which corresponds to $N_{{\rm f}}=N$.

In order to probe the SSB, we introduce an 
infinitesimal
SO(4)-breaking mass term
\begin{equation}
\Delta S_{{\rm b}}=\frac{N}{2}\varepsilon\sum_{\mu=1}^{4}m_{\mu}{\rm tr}\,(X_{\mu})^{2}
\label{eq:boson_action_bdeform}
\end{equation}
in the action, where 
\begin{equation}
m_{1}<m_{2}<m_{3}<m_{4}\ ,\label{eq:boson_mass_split}
\end{equation}
and define the order parameters for the SSB by the expectation values of 
\begin{alignat}{1}
\lambda_{\mu}=\frac{1}{N}{\rm tr}\,(X_{\mu})^{2}\ ,\label{lambda-def}
\end{alignat}
where no sum over $\mu$ is taken. 
Due to the ordering (\ref{eq:boson_mass_split}),
the expectation values obey 
\begin{alignat}{1}
\langle\lambda_{1}\rangle>\langle\lambda_{2}\rangle>\langle\lambda_{3}\rangle>\langle\lambda_{4}\rangle\label{lambda-ordering}
\end{alignat}
at finite $\varepsilon$.
If the expectation
values $\langle\lambda_{\mu}\rangle$ ($\mu=1,\cdots,4$) does not
approach the same value
in the $\varepsilon\rightarrow 0$ limit after
taking the large-$N$ limit,
we can conclude that the SSB occurs.

This issue has been studied in ref.~\cite{Nishimura:2004ts},
where explicit calculations based on the GEM
were carried out assuming that the SO(4) symmetry is broken down either
to SO(2) or to SO(3).  
For $r=1$, 
the order parameters turn out be 
\begin{alignat}{1}
\left\langle \lambda_{1}\right\rangle 
=\left\langle \lambda_{2}\right\rangle \sim2.1\ ,
\quad\left\langle \lambda_{3}\right\rangle \sim1.0\ ,
\quad\left\langle \lambda_{4}\right\rangle \sim0.8
\quad\quad & \mbox{for the \ensuremath{{\rm SO(2)}} vacuum}\ ,
\label{eq:previous_result}\\
\left\langle \lambda_{1}\right\rangle =
\left\langle \lambda_{2}\right\rangle =
\left\langle \lambda_{3}\right\rangle \sim1.75\ ,
\quad\left\langle \lambda_{4}\right\rangle \sim0.75
\quad\quad & \mbox{for the \ensuremath{{\rm SO(3)}} vacuum}\ .
\label{eq:previous_result_SO3}
\end{alignat}
The free energy was calculated in each vacuum,
and the SO(2)-symmetric vacuum was found to
have a lower value.

\section{Application of the CLM to the SO(4)-symmetric matrix model \label{sec:Application-of-the}}

In this section, we describe how we apply the CLM to the SO(4)-symmetric
matrix model (\ref{part-fn-with-det}). 
Let us rewrite the partition function as 
\begin{alignat}{1}
Z & =\int dX\,w(X)\ ,
\quad\quad w(X)=
\left(\det D\right)^{N_{{\rm f}}}
e^{-(S_{{\rm b}}+\Delta S_{{\rm b}})}\ ,
\label{part-fn-with-det-rewrite}
\end{alignat}
including the symmetry breaking term (\ref{eq:boson_action_bdeform}).
Then, the drift term that appears in the Langevin equation
is given by 
\begin{alignat}{1}
(v_{\mu})_{ij} & =
\frac{\partial\ln w(X)}{\partial(X_{\mu})_{ji}}
=-N\,\left(1+\varepsilon m_{\mu}\right)
(X_{\mu})_{ij}+N_{{\rm f}}\,
(D^{-1})_{i\alpha,j\beta}(\Gamma_{\mu})_{\beta\alpha} \ ,
\label{drift-term}
\end{alignat}
as a function of the Hermitian matrices $X_{\mu}$. 
Note that the second term in (\ref{drift-term}) is not Hermitian 
in general corresponding
to the fact that the fermion determinant is complex. 
Accordingly, the definition of $X_{\mu}$ and the drift term (\ref{drift-term})
are extended to general complex matrices
\footnote{In order to keep the matrices $X_{\mu}^{(\eta)}(t)$ close 
to Hermitian, we use the idea of gauge cooling 
proposed in ref.~\cite{Seiler:2012wz}.
Here we define the Hermiticity norm 
${\cal N}_{{\rm H}}=\frac{1}{4N}
\sum_{\mu=1}^{4}{\rm tr}
\left[\left(X_{\mu}-X_{\mu}^{\dagger}\right)
\left(X_{\mu}-X_{\mu}^{\dagger}\right)^{\dagger}\right]\ ,$
which measures the deviation of $X_{\mu}$ from a Hermitian configuration,
and minimize this norm using an ${\rm SL}(N,\mathbb{C})$ transformation
after each Langevin step.}. 

Hereafter, the parameters $m_{\mu}$ in the SO(4)-breaking
term (\ref{eq:boson_action_bdeform}) are chosen as 
\begin{equation}
(m_{1},m_{2},m_{3},m_{4})=(1,2,4,8)\ ,
\label{eq:boson_mass_used}
\end{equation}
and the Langevin step-size is chosen as $\Delta t=2.0\times10^{-4}$.

\begin{figure}[t]
\centering{}
\includegraphics[width=7cm]{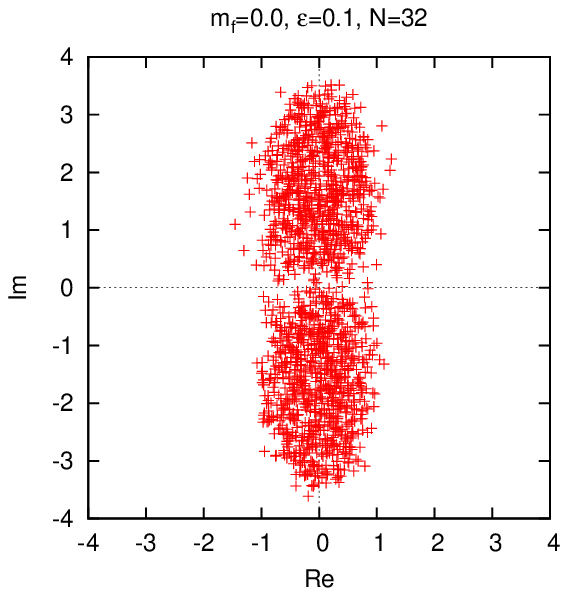}
\includegraphics[width=7cm]{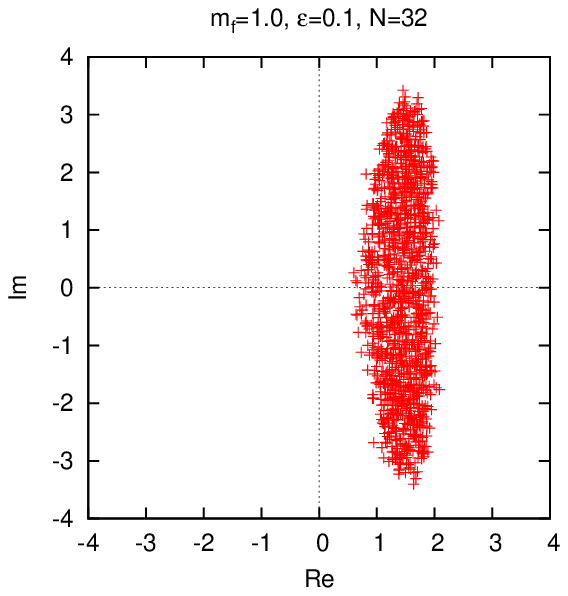}
\caption{The scatter plot for the eigenvalues of the Dirac operator obtained
during the complex Langevin simulation of 
the undeformed model (\protect\ref{part-fn-with-det-rewrite}) 
for $\varepsilon=0.1$ with $N=32$ (Left) and that of the deformed
model defined by (\protect\ref{part-fn-with-det-rewrite-deformed}) 
and (\protect\ref{eq:4-th_mass}) 
for $\varepsilon=0.1$ with $m_{{\rm f}}=1.0$ and $N=32$ (Right).
\label{fig:ev-distribut}}
\end{figure}

Let us now discuss the singular-drift problem, 
which is associated with the eigenvalues of the Dirac operator $D$ close
to zero. In Fig.~\ref{fig:ev-distribut} (Left), we plot the eigenvalue
distribution of the Dirac operator obtained during the complex Langevin
simulation for $\varepsilon=0.1$ with $N=32$. We find that there
are many eigenvalues close to zero, which suggests that the singular-drift
problem occurs.
This problem actually occurs for $\varepsilon\leq0.5$,
which makes the extrapolation to $\varepsilon=0$ difficult.

In order to avoid this problem, we add a fermion bilinear term 
which deforms the Dirac operator (\ref{eq:dirac_op}),
so that the partition function of the deformed model is defined as 
\begin{alignat}{1}
 & \tilde{Z}=\int dX\,\tilde{w}(X)\ ,
\quad\quad\tilde{w}(X)=\left(\det\tilde{D}\right)^{N_{{\rm f}}}
e^{-(S_{{\rm b}}+\Delta S_{{\rm b}})}\ ,\nonumber \\
 & \tilde{D}_{i\alpha,j\beta}=
\sum_{\mu=1}^{4}(\Gamma_{\mu})_{\alpha\beta}
\Big((X_{\mu})_{ij}+M_{\mu}\delta_{ij}\Big)\ .
\label{part-fn-with-det-rewrite-deformed}
\end{alignat}
Note that the deformation explicitly 
breaks the SO(4) symmetry of the original model (\ref{part-fn-with-det}).
Here we choose the parameters $M_{\mu}$ in such a way that the SO(4)
symmetry is broken minimally. 
Taking account of the ordering (\ref{lambda-ordering}),
we can preserve an SO(3) symmetry at $\varepsilon=0$ by choosing
\begin{equation}
M_{\mu}=\left(0,0,0,m_{{\rm f}}\right)\ .\label{eq:4-th_mass}
\end{equation}
We can then ask whether the SO(3) symmetry of this deformed model
is spontaneously broken in the large-$N$ limit.

In Fig.~\ref{fig:ev-distribut} (Right), we plot the eigenvalue distribution
of the Dirac operator (\ref{part-fn-with-det-rewrite-deformed}) 
obtained during the complex Langevin simulation of the deformed model
for $\varepsilon=0.1$ with $m_{{\rm f}}=1.0$ and $N=32$.
We find that the distribution is shifted in the real direction,
which is understandable since, at large $m_{{\rm f}}$, the eigenvalue
distribution of the Dirac operator would be distributed around $m_{{\rm f}}$.
As a result, the distribution avoids the singularity even for $\varepsilon=0.1$
in contrast to the undeformed ($m_{{\rm f}}=0$) case. Therefore,
we can extrapolate $\varepsilon$ to zero using data obtained with
smaller $\varepsilon$ for finite $m_{{\rm f}}$. 
Eventually, we extrapolate
the deformation parameter $m_{{\rm f}}$ to zero, and compare the
results with the prediction (\ref{eq:previous_result}) obtained by
the GEM for the original model.

\section{Results of our analysis \label{sec:results}}

Let us recall that we have introduced an O($\varepsilon$) mass term
(\ref{eq:boson_action_bdeform}) for the bosonic matrices, which breaks
the SO(4) symmetry explicitly. In order to probe the SSB, we need
to take the large-$N$ limit with fixed $\varepsilon$, and then make
an extrapolation to $\varepsilon=0$ afterwards.
In what follows, 
we present our results after taking the large-$N$ limit.

When we extrapolate $\varepsilon$ to zero,
it is convenient to consider the ratio 
\begin{alignat}{1}
\rho_{\mu}(\varepsilon,m_{{\rm f}})=
\frac{\langle\lambda_{\mu}\rangle_{\varepsilon,m_{{\rm f}}}}
{\sum_{\nu=1}^{4}\langle\lambda_{\nu}\rangle_{\varepsilon,m_{{\rm f}}}}\ .
\label{rho-def}
\end{alignat}
This is motivated from the fact that 
the mass term (\ref{eq:boson_action_bdeform})
tends to make all the expectation values 
$\langle\lambda_{\mu}\rangle_{\varepsilon,m_{{\rm f}}}$
smaller than the value to be obtained in the $\varepsilon\rightarrow0$
limit. 
By taking the ratio (\ref{rho-def}), the finite $\varepsilon$
effects are canceled by the denominator, and the extrapolation to
$\varepsilon=0$ becomes easier. 

\begin{figure}[t]
\centering{}
\includegraphics[width=7cm]{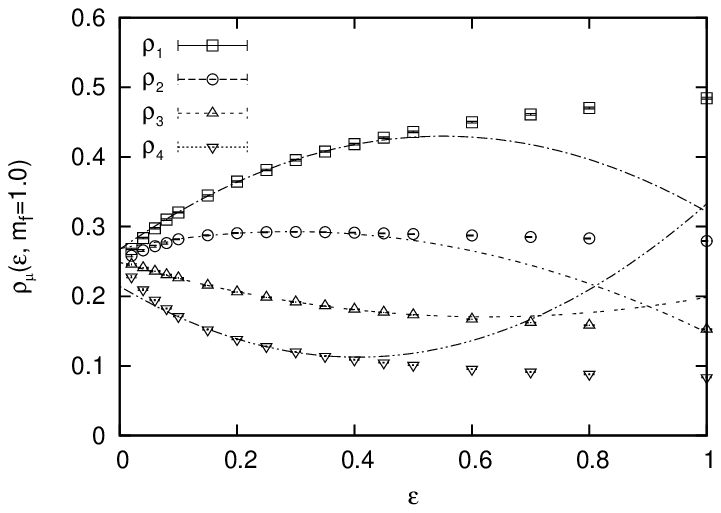}
\includegraphics[width=7cm]{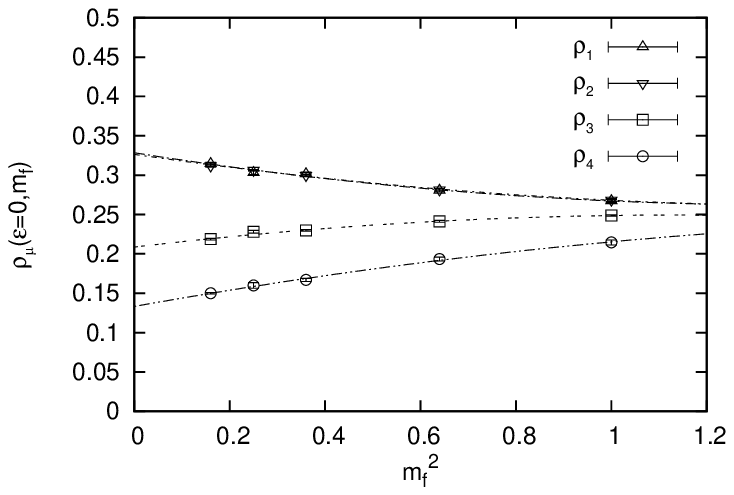} 
\caption{(Left) The ratios $\rho_{\mu}(\varepsilon,m_{{\rm f}})$ obtained
after taking the large-$N$ limit for the deformed model defined by
(\protect\ref{part-fn-with-det-rewrite-deformed}) and (\protect\ref{eq:4-th_mass})
are plotted against $\varepsilon$ for $m_{{\rm f}}=1.0$. The lines
represent fits to the quadratic form $a+b\varepsilon+c\varepsilon^{2}$.
(Right) The extrapolated values $\lim_{\varepsilon\rightarrow0}\rho_{\mu}(\varepsilon,m_{{\rm f}})$
for the deformed model defined by (\protect\ref{part-fn-with-det-rewrite-deformed})
and (\protect\ref{eq:4-th_mass}) 
are plotted against $m_{{\rm f}}^{2}$. The lines represent fits 
to the quadratic form $a+bx+cx^{2}$ with $x=m_{{\rm f}}^{2}$ using
the data within the region $0.4\leq m_{{\rm f}}\leq1.0$.
\label{fig:extrolate_eps}}
\end{figure}

In Fig.~\ref{fig:extrolate_eps} (Left), we plot the ratio (\ref{rho-def})
against $\varepsilon$ for $m_{{\rm f}}=1.0$.
Using the criterion proposed in ref.~\cite{Nagata:2016vkn}, we can decide
which data points in the small $\epsilon$ region 
suffer from the singular drift problem and hence
should be excluded from the extrapolation to $\epsilon=0$.
We find that the fitting curves for 
$\rho_{1}(\varepsilon,m_{{\rm f}})$ and 
$\rho_{2}(\varepsilon,m_{{\rm f}})$
approach the same value at $\epsilon=0$,
while the others approach smaller values.
This implies that the SSB from SO(3) to SO(2) occurs 
in the deformed model.

In Fig.~\ref{fig:extrolate_eps} (Right), we plot the extrapolated
values $\lim_{\varepsilon\rightarrow0}\rho_{\mu}(\varepsilon,m_{{\rm f}})$
thus obtained against $m_{{\rm f}}^{2}$. We find that our
results within $0.4\le m_{{\rm f}}\le1.0$ can be nicely fitted to
the quadratic behavior, 
which is motivated by a power series 
expansion\footnote{The odd order terms in $m_{{\rm f}}$ 
do not appear due to the symmetry
$m_{{\rm f}}\rightarrow-m_{{\rm f}}$ of the expectation values.}
of the expectation
values $\langle\lambda_{\mu}\rangle_{\varepsilon,m_{{\rm f}}}$ with
respect to $m_{{\rm f}}$.
Extrapolating $m_{{\rm f}}$ to zero, 
we obtain $\lim_{m_{{\rm f}}\rightarrow0}
\lim_{\varepsilon\rightarrow0}\rho_{\mu}(\varepsilon,m_{{\rm f}})=0.328(4),$
0.326(2), 0.208(2), 0.133(2) for $\mu=1,2,3,4$, which shows that
the SO(4) symmetry of the undeformed model ($m_{{\rm f}}=0$) 
is spontaneously
broken down to SO(2). 
Moreover, using an exact result 
$\sum_{\mu=1}^{4}\langle\lambda_{\mu}\rangle=4+2r=6$
\cite{Nishimura:2001sq} for the present $r=1$ case, 
we obtain 
\begin{equation}
\left\langle \lambda_{1}\right\rangle =1.97(2)\ ,
\quad\left\langle \lambda_{2}\right\rangle =1.96(1)\ ,
\quad\left\langle \lambda_{3}\right\rangle =1.25(1)\ ,
\quad\left\langle \lambda_{4}\right\rangle =0.80(1)\ ,
\label{eq:deform-4th_result}
\end{equation}
which agree well with the results (\ref{eq:previous_result}) obtained
by the GEM. Here we emphasize that in the GEM,
the true vacuum was determined by comparing the free energy obtained
for the SO($2$) vacuum and the SO($3$) vacuum. In contrast, the
CLM enables us to determine the true vacuum directly without having
to compare the free energy for different vacua. 

\section{Summary and discussion \label{sec:Summary-and-discussion} }

We have shown that the CLM can be successfully applied
to a matrix model, in which the SSB of SO(4) is expected to occur due to
the phase of the complex fermion determinant. 
For this success,
it was crucial to overcome the singular-drift problem associated with
the appearance of nearly zero eigenvalues of the Dirac operator. 
Our strategy was to deform the Dirac operator
in such a way that the singular-drift problem is avoided while maintaining
the qualitative feature of the vacuum as much as possible.
Extrapolating the deformation parameter, we were able 
to obtain the order parameters, which turned out
to be consistent with the prediction by the GEM. 

The CLM with the proposed strategy can be directly applied to the
type IIB matrix model, which is conjectured to be a nonperturbative
formulation of type IIB superstring theory 
in ten dimensions \cite{Ishibashi:1996xs}.
While the SO(10) symmetry of the model is expected to be spontaneously
broken down to SO(4) for consistency with our 4d space-time, the GEM
predicts that it 
is spontaneously broken down to SO(3) rather than SO(4). 
It would be interesting to investigate this issue using the CLM extending
the present work.

We consider that the same strategy would enable the 
application of
the CLM to finite density QCD at low temperature and high density,
where various exotic condensates are speculated to form
due to the complex fermion determinant. 
We can try to find some deformations
which allow us to extrapolate the deformation parameter to zero
within the region of validity.
Here the new criterion \cite{Nagata:2016vkn}
for justifying the CLM should be useful, as it is the case
in the present work \cite{Ito:2016efb}.

\section*{Acknowledgements }


The authors would like to 
thank K.N.~Anagnostopoulos, T.~Azuma, K.~Nagata, S.K.~Papadoudis
and S.~Shimasaki for valuable discussions. Y.~I.\ is supported
by JICFuS. J.~N.\ is supported in part by Grant-in-Aid for Scientific
Research (No.\ 23244057 and 16H03988) 
from Japan Society for the Promotion of Science. 

\end{document}